\documentclass{article}
\usepackage{graphicx}

\begin{document}

\begin{center}
{\bf \LARGE A two-wave dynamo model by Zharkova et al. (2015) disagrees with data on long-term solar variability}
\end{center}

I.G. Usoskin$^{1,2}$, G.A. Kovaltsov$^3$,

$^1$ReSoLVE Centre of Excellence, University of Oulu, Finland

$^2$Sodankyl\"a Geophysical Observatory (Oulu unit), University of Oulu, Finland

$^3$Ioffe Physical-Technical Institute, St.Petersburg, Russia

\abstract{
A two-wave dynamo model was recently proposed by Zharkova et al. (2015, Zh15 henceforth), which aims at
 long-term predictions of solar activity for millennia ahead and backwards.
Here we confront the backward predictions for the last 800 years with known variability of
 solar activity, using both direct sunspot observations since 1610 and reconstructions based
 on cosmogenic nuclide data.
We show that the Zh15 model fails to reproduce the well-established features of the solar activity
 evolution during the last millennium.
This means that the predictive part for the future is not reliable either.
}

\section{Introduction}

A new view on the dynamo mechanism operated in the Sun's convection zone was proposed recently
 by Zharkova et al. (\cite{zharkova15} Zh15, henceforth).
According to Zh15, the model is able to predict variability of solar activity on millennial time scale.
In particular, in Fig.3 of Zh15, a ``prediction'' of solar variability is shown for 1200 years ahead and 800 years
 backwards.
While future predictions are unverifiable, the past solar activity is known quite well for the last millennium
 and can be easily confronted with the predictions by Zh15.
As we show here, Zh15 work fails in reconstruction of the past solar activity and accordingly is not trustworthy in predictions.

\section{Comparison with data}

We confronted the results of the Zh15 ``prediction'' of the past solar activity with available data obtained either directly
 from sunspot observations for the last 400 years or from cosmogenic radionuclide ($^{14}$C in tree rings and $^{10}$Be in ice cores)
 data, which form a direct proxy for
 cosmic rays variability and thus for solar magnetic activity \cite{beer12,usoskin_LR_13}.
Cosmogenic data, particularly radiocarbon $^{14}$C, cannot reproduce the 11-year cycle.
Since we focus here on the centennial variability, we further discuss decadal data which is the time resolution of many
 cosmogenic nuclide series (e.g., \cite{reimer_09}).
Accordingly, the data from Zh15 and all other data with sub-decadal time resolution were resampled to become 10-year averages.
For the ZH15 data\footnote{Since the authors of Zh15 have strictly rejected to provide original data for an independent test, 
 we extracted their data from
 Figure 3 of Zh15 paper, which may lead to some minor inexactitudes but does not affect the centennial pattern.}
  we used the modulus of the ``summary wave'' of the Sun's poloidal magnetic field shown in
 their Figure 3.

\begin{figure}[t]
\centering \resizebox{10cm}{!}{\includegraphics{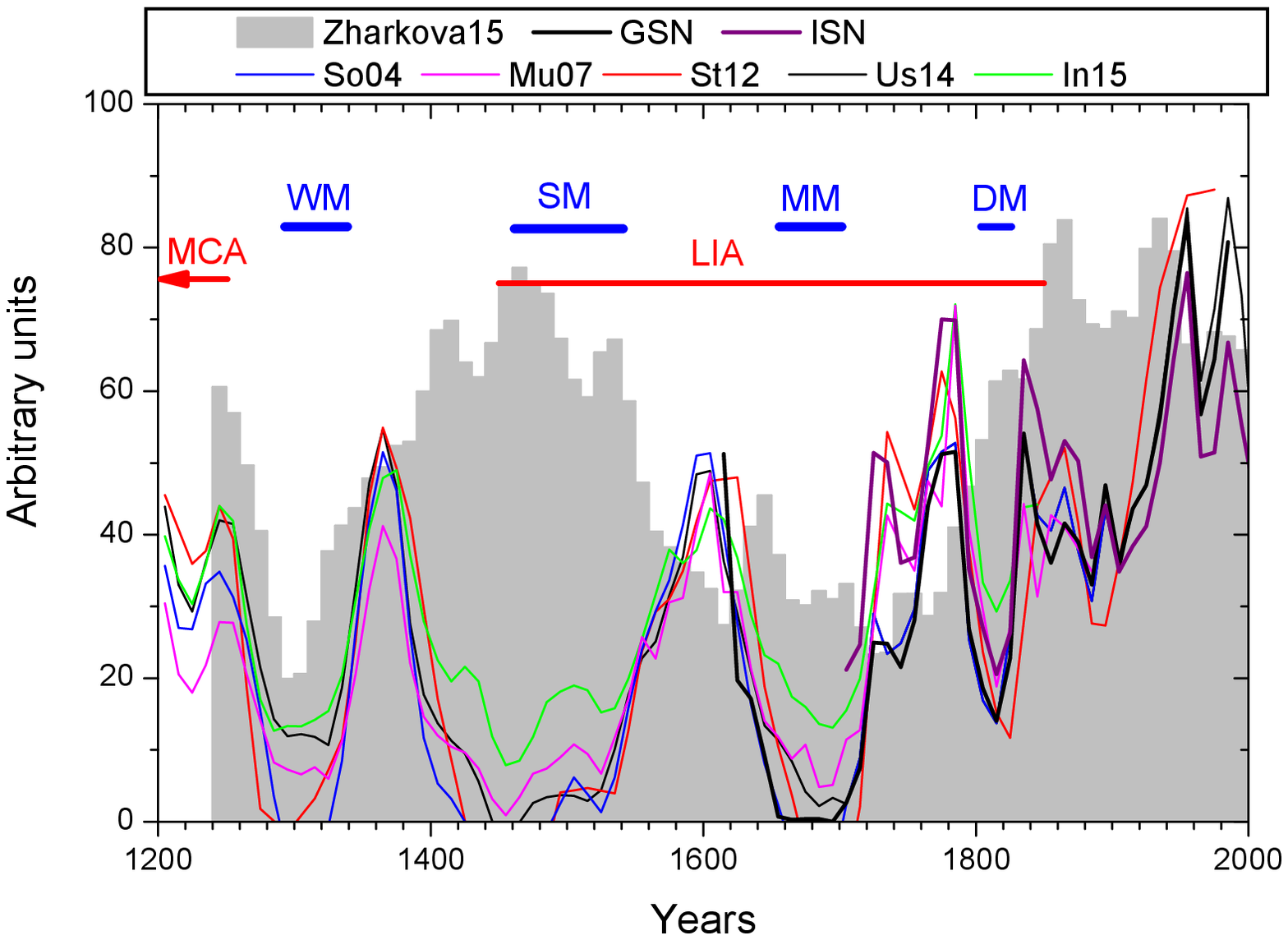}}
\caption{
Deacadal (10-yr averaged) solar variability.
The grey bars represent the modulus of the ``summary wave'' from Figure 3 of Zh15 (see footnote 1), while colored curves depict
 different measured and proxy data for solar variability for the period 1200\,--\,2000 AD, as denoted in the legend:
 GSN -- group sunspot number \cite{hoyt98}; ISN -- international sunspot number, v.2.0 \cite{clette14};
 So14 -- sunspot reconstruction from $^{14}$C \cite{solanki_Nat_04}; Mu07 -- modulation potential from $^{14}$C \cite{muscheler07};
 St12 -- modulation potential from $^{14}$C and $^{10}$Be \cite{steinhilber12};
 Us14 -- sunspot reconstruction from $^{14}$C \cite{usoskin_AAL_14};
 In15 -- modulation potential from $^{14}$C and $^{10}$Be \cite{inceoglu15}.
 All curves are given in arbitrary units for visibility and show a similar variability.
 The grand minima of solar activity are denoted by blue horizontal bars:
   Wolf (WM), Sp\"orer (SW), Maunder (MM) and Dalton (DM) minima (see dates in Table~\ref{Tab}).
 Periods of the Medieval Climate Anomaly (MCA) and the Little Ace Age (LIA) \cite{IPCC_AR5} are denoted by red horizontal bars.
 }
\label{Fig:SN}
\end{figure}
The comparison is shown in Figure~\ref{Fig:SN}.
One can see that the predicted curve by Zh15 (grey bars) disagrees with the real solar variability (colored curves).
For example, the linear Pearson's correlation between the result by Zh15 and that by \cite{steinhilber12} (St12) curve is $0.14\pm 0.26$,
 which means no correlation.
Other long-term curves also imply no or insignificant correlation.

The Wolf minimum ca. 1300 is correctly reproduced by the model but it is the only success.
Solar activity is predicted to be very high, comparable to the modern grand maximum, during the period of 1350\,--\,1650,
 but in fact the longest and deepest minimum during the last millennium, the Sp\"orer minimum, took place during 1390\,--\,1550
 (see Table~\ref{Tab}).
Thus, the Sp\"orer minimum is not reproduce at all, on the contrary, a high activity is predicted by the Zh15 model.
A suppression centered at the Maunder minimum is observed in the model prediction but it is much longer (about 200 years)
 and disagrees with the periods of high activity ca. 1600 and in the second half of the 18th century \cite{inceoglu15,stuiver80}.
The Dalton minimum is not reproduced correctly, as it falls on the period of the fast growth of the predicted activity.
The Gleissberg (or Modern) minimum of activity ca. 1900 also disagrees with the predicted data, which reach the centennial maximum
 at that time.

Thus, we conclude that the model of Zh15 fails in reconstructing solar activity in the past, especially during the
 14th through 16th centuries when the prediction is in antiphase with the real solar activity.
This makes long-term predictions based on this model unreliable.

\section{Time mismatch}

Zh15 paper claims that the model prediction ``corresponds very closely to the sunspot data observed in the past 400...
 [and] predicts correctly many features from the past, such as: 1) an increase in solar activity during the Medieval Warm period;
 2) a clear decrease in the activity during the Little Ice Age, the Maunder Minimum and the Dalton Minimum; 3) an increase in
 solar activity during a modern maximum in 20th century.''
Here we show that this statement, based on a visual inspection of Figure 3 of Zh15, is not correct.
We noticed that most of the horizontal bars indicating different periods in Figure 3 of Zh15 paper are misleading.
Since references to these bars are not provided, we copied the dates directly from Figure 3 and collected them into Table~\ref{Tab} here
 along with the appropriate dates.
\begin{table}
\caption{Specific periods during the last 800 years with the dates according to Figure 3 of Zh15 and the
 appropriate dates, and references for the latter.}
\begin{tabular}{llll}
\hline
Period & Dates Zh15 & Appropriate dates & Reference\\
\hline
Modern maximum & 1890\,--\,2000 & 1940\,--\,2009 & \cite{solanki_Nat_04}\\
Dalton minimum & 1705\,--\,1800 & 1790\,--\,1820 & \cite{usoskin_LR_13}\\
Maunder minimum & 1590\,--\,1680 & 1645\,--\,1715 & \cite{eddy76,stuiver80}\\
Sp\"orer minimum & does not exist & 1390\,--\,1550 & \cite{usoskin_LR_13,inceoglu15,stuiver80}\\
Wolf minimum & not shown & 1270\,--\,1340 & \cite{usoskin_LR_13,inceoglu15,stuiver80}\\
Little Ice Age & 1560\,--\,1765 & 1450\,--\,1850 & \cite{IPCC_AR5}$^\dagger$\\
Medieval Warm Period & 1350\,--\,1530 & 950\,--\,1250 & \cite{IPCC_AR5}$^\dagger$\\
\hline
$^\dagger$ see Table 5.1.\\
\end{tabular}
\label{Tab}
\end{table}

One can see that, for example the dates of the Medieval Warm Period (conventionally called the Medieval
 Climate Anomaly) was shifted by Zh15 for several hundred years ahead.
The appropriate dates of the Anomaly is beyond the frame of Figure 3 in Zh15.
Other dates are also misplaced to `match' the prediction curve.

\section{Conclusion}

As shown here, the long-term reconstruction of solar activity, using the two-wave dynamo model, as presented by Zh15
 fails in reproducing the well-established pattern of solar variability during the last millennium.
Thus this model cannot be trusted for the future predictions.

\bibliographystyle{unsrt}

\end{document}